\documentclass[onecolumn,prd,amsmath,amsssymb,groupedaddress,11pt]{revtex4}

\usepackage{graphicx}
\usepackage{dcolumn}
\usepackage{bm}
\usepackage{amssymb}
\usepackage{amsmath}
\def\lsim{\mathrel{\mathstrut\smash{\ooalign{\raise2.5pt\hbox{$<$}\cr\lower2.5pt\hbox{$\sim$}}}}}
\def\gsim{\mathrel{\mathstrut\smash{\ooalign{\raise2.5pt\hbox{$>$}\cr\lower2.5pt\hbox{$\sim$}}}}}
\def\ba{\begin{eqnarray}}
\def\ea{\end{eqnarray}}
\def\be{\begin{equation}}
\def\ee{\end{equation}}

\def\beq{\begin{eqnarray}}
\def\eeq{\end{eqnarray}}

\def\lsim{\mathrel{\rlap{\lower3pt\hbox{\hskip0pt$\sim$}}
    \raise1pt\hbox{$<$}}}         
\def\gsim{\mathrel{\rlap{\lower4pt\hbox{\hskip1pt$\sim$}}
    \raise1pt\hbox{$>$}}}         

\def\bea{\begin{eqnarray}}
\def\eea{\end{eqnarray}}
\def\({\left(}
\def\){\right)}
\def\ie{{\it i.e. }}
\def\nn{\nonumber}

\begin{document}

\title{Cascading Gravity and Degravitation}

\author{Claudia de Rham$^{1,\, 2}$, Stefan Hofmann$^{1,\, 3}$, Justin Khoury$^{1}$, Andrew J. Tolley$^{1}$}

\affiliation{
\nobreak{$^1$Perimeter Institute for Theoretical Physics, 31 Caroline St. N., Waterloo, ON, N2L 2Y5, Canada} \\
\nobreak{$^2$ Dept. of Physics \& Astronomy, McMaster University, Hamilton ON, Canada}\\
\nobreak{$^3$ NORDITA, Roslagstullsbacken 23, 106 91 Stockholm, Sweden}}

\begin{abstract}
\begin{center}
{\bf Abstract}
\end{center}
\noindent
We construct a cascading brane model of gravity in which
the behavior of the gravitational force law interpolates from $(n+4)$-dimensional to $(n+3)$-dimensional all the way down to 4-dimensional from
longer to shorter length scales. We show that at the linearized
level, this model exhibits the features necessary for degravitation
of the cosmological constant. The model is shown to be ghost free
with the addition of suitable brane kinetic operators, and we demonstrate this using a number of
independent procedures. Consequently this is a
consistent IR modification of gravity, providing a promising
framework for a dynamical, degravitating solution of the
cosmological constant problem.
\end{abstract}
\maketitle
\newpage
\setcounter{tocdepth}{3}
\tableofcontents

\section{Introduction}

The remarkable theoretical challenge posed by the cosmological constant problem \cite{weinberg} has spurred many attempts at directly modifying Einstein's gravity at large distances.
A compelling example of one such infrared (IR) modification is the DGP brane-world model~\cite{DGP,gregorygia}.
In this scenario, our visible world is confined to a brane in an infinite 5D bulk.
The inclusion of a graviton kinetic term on the brane recovers the usual gravitational force law scaling, $1/r^2$, at short distances, but at large distances it asymptotes to the 5D scaling, $1/r^3$.

The resulting weakening of gravity on large scales can have profound implications for the cosmological constant
problem~\cite{dgs}. In particular, these models have in common the feature that the 4$D$ graviton acquires a mass, or is a
resonance --- an infinite superposition of massive states.
In a recent proposal~\cite{degrav}, two of us put forward the idea, based on earlier work in~\cite{addg},
that if gravity is sufficiently weakened in the infrared, then vacuum energy could
effectively decouple from gravity or degravitate
over time. This would provide a causal, dynamical solution to the cosmological constant problem.
However, the degravitation phenomenon is not realized in the standard DGP model because the weakening of gravity is not sufficiently steep.
In~\cite{degrav} it was shown that degravitation requires that the 4$D$ graviton propagator grows slower than $1/p$ in the IR~\cite{degrav}, whereas in standard DGP it grows exactly
as $1/p$. An explicit model with the desired behavior for the propagator was not given in~\cite{degrav}, and the goal of the present work is to construct specific non-linear theories that can exhibit the degravitation phenomenon.

A promising way to realize degravitation is to consider higher-codimension DGP brane-world models, {\it i.e.} higher-dimensional bulks. If the bulk has six or more space-time dimensions, then the gravitational force law falls off faster in the IR
than in the standard DGP model, and as we shall see in what follows this fall off is sufficiently fast to exhibit degravitation at the linearized level. However, higher-codimension DGP scenarios have proven notoriously difficult to realize consistently for two reasons: Firstly, the 4$D$ propagator
diverges when evaluated on the brane and must be regularized~\cite{geroch,gw,claudia}. Secondly, the
simplest constructions are plagued by ghost-like instabilities~\cite{sergei,gigashif}.

In this paper we argue that both pathologies --- divergent propagator and ghost instabilities --- are resolved in the recently proposed Cascading DGP model~\cite{oriol}.
The idea is that our visible 3-brane lies within a succession of higher-dimensional branes, each with their own induced gravity terms,
embedded in one another in a flat $D$-dimensional bulk. For instance, in the simplest codimension-two case, our 3-brane is embedded in a 4-brane within a flat 6$D$ bulk.
The gravitational force law therefore ``cascades" from 4$D$ ($1/r^2$) to 5$D$ ($1/r^3$) to 6$D$ ($1/r^4$) etc., as we probe larger distances on
the 3-brane (a similar cascading behavior of the force law was also obtained recently in a different codimension-two framework~\cite{Kaloper:2007ap}, and closely related work on intersecting branes was discussed in \cite{Corradini:2007cz} with somewhat different motivations).

In Sec.~\ref{degravrev} we review the idea of degravitation to motivate the study of higher-codimension brane set-ups, and explain how these satisfy the properties expected of the linear theory using the spectral representation of the propagator, namely that the 4$D$ propagator in momentum space grows more slowly than $1/p$ as $p\rightarrow 0$.
Cascading DGP gravity, therefore, provides a candidate nonlinear realization of the degravitation idea
and opens the door to a dynamical resolution to the cosmological constant problem.

We begin with a scalar analogue in Sec.~\ref{scalar}, consisting of a bulk scalar field with induced kinetic terms on each of the branes.
We find that the usual divergence of the 4$D$ propagator is cured by the presence of the higher-dimensional branes ---
the extra kinetic terms act as regulators for the 3-brane propagator.
This framework is then generalized to gravity in Sec.~\ref{gravSing}, where we find that the cascading set-up similarly regulates the usual divergences, however it does not automatically avoid the ghost instability characteristic of higher-codimension
DGP models~\cite{sergei,gigashif}. As in the pure codimension-$n$ DGP model, adding an $R_4$ term on the 3-brane
results in a ghost instability. This can readily be seen at the level of the one-particle exchange amplitude, which naturally
splits into a transverse-traceless (or massive graviton) contribution and a scalar component. In the IR, these combine to give the usual tensor structure of
a massless graviton in $D$ dimensions, as it should. In the UV, however, the scalar contribution is negative, signaling a ghost, {\it i.e.} the scalar part of the Green's function
does not allow a well-defined spectral representation.

In~\cite{oriol} it was argued in the codimension-two case that the ghost is cured by including a sufficiently large tension on the (flat) 3-brane.
Here we follow a different approach by noting that an $R_4$ term is not the most general induced gravity action on a 3-brane.
In Sec.~\ref{codnreg} we present two alternatives, obtained through regularization of the codimension-$n$ brane. An important concern when
smearing out branes is maintaining gauge invariance, an issue we take great care in
by utilising the St{\"u}ckelberg formalism. The approach of Sec.~\ref{codnreg} is not restricted to Cascading DGP, but applies more generally to arbitrary
codimension-$n$ DGP models.

\begin{itemize}

\item The first method (Sec.~\ref{sphreg}) consists of blowing up the 3-brane into a codimension-one sphere of radius $\Delta$. The key observation is that,
once the defect has finite thickness, its worldvolume has more than 3+1 dimensions. Consequently, it no longer makes sense to include an $R_4$ term covariantly.
Rather, the natural choice is to add an $R_{3+n}$ term, the Einstein-Hilbert term appropriate for the dimensionality of the regularized codimension-one brane.

\item The second method (Sec.~\ref{mediumstory}) promotes the 3-brane into a full $(3+n)$-dimensional object of radius $\Delta$. The natural choice in this case is to include an $R_D$ term within its worldvolume, albeit with a different Newton's constant than that in the bulk. Thus the codimension-$n$ is understood as a medium with non-zero gravitational permeability, as first introduced in~\cite{massimo}.

\end{itemize}

\noindent Remarkably, these additional kinetic terms on the regularized brane have dramatic consequences for the scalar degrees of freedom: Both methods yield the same, ghost-free exchange amplitude, displaying the tensor structure characteristic of the $D$-dimensional massless graviton on all scales. The effect of these terms is very similar to those considered in \cite{gigashif} where it was shown that adding an $R_6$ kinetic term on a 3-brane in 6D would, for sufficiently large value of this term, remove the ghost. Another way to understand these operators is that one is adding extrinsic curvature terms into the codimension-two brane action. While these would normally render the boundary value problem ill-defined without the addition of higher derivative terms in the bulk, here we can make sense of these terms by suitably regularizing the brane.

That these two methods yield the same answer can be understood most transparently in the language of effective field theory, as shown in Sec.~\ref{eff}. In the thin-brane limit, $\Delta\rightarrow 0$, we can perform a Kaluza-Klein reduction within the worldvolume of our defect to obtain an effective action on the 3-brane. Note that dimensional reduction is performed only within the defect, while the bulk remains infinite in extent. The resulting effective action reduces to an $R_4$ term, although with additional scalar-tensor couplings to the extra-dimensional moduli. These scalar-tensor couplings are essential in modifying the scalar part of the 4$D$ exchange amplitude, thereby rendering it ghost-free.

As mentioned earlier, these regularization schemes apply quite generally to any higher-codimension DGP model. In Sec.~\ref{gravreg}, however, we focus on their implications
for cascading DGP. In the codimension-two case, in particular, the one-particle exchange amplitude on the 3-brane interpolates between the usual $1/r$ at short distances and
$1/r^3$ at large distances, while the tensor structure remains that of a 6D massless graviton on all scales.

The decoupling limit~\cite{lpr} of the codimension-two cascading theory offers yet another perspective on how the ghost is excised from the spectrum, as shown in Sec.~\ref{decoupling}.
By taking $M_5,M_6\rightarrow \infty$ keeping the strong coupling scale $\Lambda_{\rm s} = \left(M_6^{16}/M_5^9\right)^{1/7}$ fixed, we can integrate out the 6D bulk physics to
obtain a local, five-dimensional effective theory for the 5D metric and the scalar (brane-bending) mode $\pi$~\cite{oriol}. For the usual choice of $R_4$ term on
the 3-brane, the $\pi$ scalar ends up having a negative kinetic term on the 3-brane --- hence it is a ghost. On the other hand, the 3-brane effective action
mentioned above contributes a positive induced kinetic term for $\pi$, rendering this mode healthy. In fact, the decoupling analysis shows that our effective action
is part of a broader class of 3-brane lagrangians involving $\pi$ that yield a ghost-free amplitude. The inclusion of a non-zero tension considered in~\cite{oriol} is another example.

At face value the 6$D$ tensor structure of the 4$D$ graviton propagator is inconsistent with solar system tests of General Relativity. In other words, the codimension-two Cascading
DGP framework yields $-1/4$ as the $TT'$ coefficient in the exchange amplitude, whereas General Relativity yields $-1/2$. This mismatch persists for arbitrarily small graviton mass,
a paradox known as the van Dam-Veltman-Zakharov (vDVZ) discontinuity~\cite{vDVZ}. As in the standard DGP model,  we expect this effect to be an artifact of perturbation theory~\cite{ddgv}. In the vicinity of astrophysical sources, within the so-called Vainshtein radius, the one-particle exchange approximation breaks down and the scalar modes become strongly coupled. The end result is that the scalar modes decouple, and Einstein gravity is a good approximation.

As argued in Sec.~\ref{strongcoupling}, the natural outcome in Cascading DGP is a succession of Vainstein radii for widely separated cross-over scales. In the codimension-two case, for instance, the scalar responsible for the $6D\rightarrow 5D$ transition is $\pi$. We therefore expect this scalar to decouple below the Vainshtein radius as in the standard DGP model in $6D$. Below this scale, the theory effectively reduces to $5D$ gravity coupled to a 3-brane with induced gravity term, {\it i.e.}, the standard DGP model. We can anticipate a second Vainshtein effect for the longitudinal mode of the graviton at sufficiently short distances.

To summarize, in this paper we show that the cascading DGP framework recently proposed in~\cite{oriol} is a compelling candidate for a consistent nonlinear realization of degravitation of the cosmological constant.
Motivated by independent regularization schemes, we derive a new effective action on the 3-brane, which eradicates the ghost instability characteristic of higher codimension DGP models.
In the codimension-two case, the resulting gravitational force law on the brane interpolates from 6D at the largest scales to 4D at the shortest via an intermediate 5D regime. This model
naturally generalizes to higher dimensions where there is a similar cascade from $(4+n)$-D to $(3+n)$-D down to 4D.

There are many important issues which are left for future work. First of all, it remains to show that degravitation takes place in the full non-linear theory, by studying the cosmological evolution
in the presence of large vacuum energy on the brane. The linearized analysis presented here certainly displays all the desired features, but it remains to be seen whether this is also the case at the non-linear level.
Moreover, one should check in detail that the extra scalar modes decouple near large sources due to non-linear (or Vainshtein) effects, a necessary condition for the theory to be phenomenologically viable.

\section{Review of the Degravitation Phenomenon}
\label{degravrev}

Almost all attempts in addressing the cosmological constant problem have focused on making the vacuum energy density small. Degravitation is a framework that instead allows
for a large cosmological constant, but suppresses its backreaction by making gravity exponentially weaker at long distances. In other words, the vacuum energy density is not small in this
picture, it is simply degravitated. A phenomenological modification to Einstein gravity that encapsulates this behavior is~\cite{dgs,addg}
\be
G_{\rm N}^{-1}(L^2\Box)G_{\mu\nu} = 8\pi T_{\mu\nu}\,,
\label{dg1}
\ee
where Newton's constant has been promoted to a derivative operator. This is such that gravity behaves as a high-pass filter with characteristic scale $L$:  $G_{\rm N}\rightarrow G_{\rm N}^{(0)}$ for $L^2(-\Box)\rightarrow \infty$ and $\rightarrow 0$ for $L^2(-\Box)\rightarrow 0$. However, as it stands this equation cannot be consistent. This is immediately seen, for instance, by noting that the Bianchi identity --- a direct consequence of general covariance in four dimensions --- is violated.

Indeed, a key observation made in~\cite{gd,degrav} is that any degravitating theory must reduce at the linearized level to a theory in which the graviton has a mass or is a resonance, \ie a continuum of massive states. That is, the linearized equations of motion must be of the form
\be
\label{gfpeom}
\left({\cal E}h\right)_{\mu\nu} + \frac{m^2(\Box)}{2}\left(h_{\mu\nu} - \eta_{\mu\nu} h\right)  = - T_{\mu\nu} \,,
\ee
where $\left({\cal E}h\right)_{\mu\nu} = -\Box h_{\mu\nu}/2 + \ldots$ is the linearized Einstein tensor. The above is a straightforward generalization of Fierz-Pauli massive gravity \cite{FP}, the only ghost-free theory of a (free) massive spin-2 particle, with the mass term promoted to a function of the derivative operator.
An immediate corollary is that any degravitating theory must describe extra degrees of freedom, corresponding to the extra polarizations of the massive graviton. Indeed, the metric fluctuation $h_{\mu\nu}$ propagates 5 degrees of freedom: 2 helicity-2 modes, 2 helicity-1 modes, and 1 helicity-0 mode.

Deconstructing the full metric fluctuation into its various helicity constituents offers a beautiful proof of concept for degravitation: the various helicities are honest gauge fields, \ie they transform in a nontrivial way under small gauge shifts. However, the full metric fluctuation is gauge invariant. In other words, the helicity-1 and helicity-0 components play the role of St{\"u}ckelberg fields that render the full metric fluctuation gauge invariant. Thus, $h_{\mu\nu}$ becomes an observable (at least in principle), and (\ref{gfpeom}) is an equation of motion for an observable quantity.
Hence, a sensible solution of (\ref{gfpeom}) has to be bounded. This is not the case for a massless fluctuation which, for a homogeneous source, develops an instability towards de Sitter.
In contrast, massive fluctuations respect Minkowski, \ie the mass term in (\ref{gfpeom}) guarantees that flat space is a consistent background, even for a cosmological term.
At this point we reach the conclusion that the mass term in (\ref{gfpeom}) gives rise to a dispersion for the gravitons of generalized Fierz-Pauli that effectively decouples them from any infrared source.
In other words, a high-pass filter must be at work and is activated under the spell of gauge invariance.
Indeed, the filter equation~(\ref{dg1}) was shown to arise as an effective equation for the helicity-2 (or Einsteinian) modes, once the other helicities have been integrated out~\cite{gd,degrav}.

The next question is what are the allowed forms for $m^2(\Box)$? Since vacuum energy is the longest-wavelength source, we are interested in the far IR behavior of this
mass operator. A useful parametrization in this regime is a power-law:
\be
m^2(\Box) = L^{-2(1-\alpha)}\Box^\alpha\,.
\ee
Clearly $\alpha < 1$ in order for this to represent an IR modification of gravity. Requiring that the corresponding propagator has a well-defined spectral representation, so that the theory is free of ghosts,
puts a lower bound on this parameter: $\alpha\geq 0$. Thus all degravitation theories can be classified at the linearized level by a single parameter $\alpha$ within the allowed range:
\be
0\leq \alpha < 1\,.
\ee
For instance, massive gravity corresponds to $\alpha = 0$, whereas the DGP model has $\alpha = 1/2$, as can be seen from~(\ref{DGPprop}) for instance. Moreover,
the logarithmic behavior as $p\rightarrow 0$ of our codimension-two cascading propagator, as shown in~(\ref{loglimit}), essentially corresponds to $\alpha\approx 0$.

\subsection{Spectral Representations}

We now present an argument that all cases in which the gravitational potential at large distances behaves as higher dimensional, correspond to $\alpha=0$ or $\alpha=1/2$. Associated with a given spectral representation of the propagator,
\be
{\mathcal G}(p_{\mu})= \int_{0}^{\infty} \frac{\rho(s)}{p^2+s} d s\,,
\ee
we have the `Newtonian potential'
\be
V(r)= \int_{0}^{\infty} \rho(s) \frac{\exp{(-\sqrt{s} r)}}{r} d s\,.
\ee
In $D$ space-time dimensions, the Newtonian potential scales as $V(r)
\sim 1/r^{D-3}$, and so in a codimension-$n$ DGP model, we expect
that at large distances we feel the full $(4+n)$-dimensions so that
$V(r) \sim 1/r^{1+n}$. Since the potential at large distances is
determined by the $s \rightarrow 0$ behavior of the spectral
density, it is clear that $\rho(s)$ must scale as $\rho(s) \sim
s^{n/2-1}$ as $s \rightarrow 0$.  To make this concrete, define $\rho(s) = s^{n/2-1} f(s)$,
where the only requirement is that the function $f(s)$ is finite and
nonzero at $s=0$. The Newtonian potential is then
\be
V(r)= \int_{0}^{\infty} s^{n/2-1} f(s) \frac{\exp{(-\sqrt{s} r)}}{r} d s \, ,
\ee
which can be simplified by redefining the integration variable $s=x^2/r^2$ so that
\be
V(r)= \frac{2}{r^{1+n}}\int_{0}^{\infty} x^{n-1} f(x^2/r^2) \exp{(-x)} d x \, .
\ee
On taking the limit $r \rightarrow \infty$ we have
\be
\lim_{r \rightarrow \infty} V(r)= \frac{2
f(0)}{r^{1+n}}\int_{0}^{\infty} x^{n-1} \exp{(-x)} d x=\frac{2 f(0)
\Gamma(n)}{r^{1+n}}\,,
\ee
as anticipated.

We are now in a position to show that all higher codimension cases
corresponds to $\alpha=0$. In the limit $p \rightarrow 0$, the propagator is given by
\be
{\mathcal G}(0)=\int_{0}^{\infty} \frac{s^{n/2-1} f(s)}{s} d s\,.
\ee
This integral is convergent for small $s$ provided that $n>2$, and
hence $1/{\mathcal G}(p_{\mu}) \sim p^{2\alpha}$ with $\alpha=0$ for all codimensions
$n>2$. For the case $n=2$ the integral is IR log divergent, ${\mathcal G}(p_{\mu})
\sim \log(p)$, which is consistent with the above results. For the
case $n=1$ we have
\be
{\mathcal G}_{n=1}(p_{\mu})=\int_{0}^{\infty} \frac{s^{-1/2} f(s)}{p^2+s} d s\,.
\ee
By redefining $s=p^2 y$, we find
\be
\lim_{p \rightarrow 0} {\mathcal G}_{N=1}(p_\mu)=\lim_{p \rightarrow 0} p^{-1}
\int_{0}^{\infty} \frac{y^{-1/2} f(p^2 y)}{1+y} d y
=p^{-1}f(0)\int_{0}^{\infty} \frac{y^{-1/2}}{1+y} d y =p^{-1} \pi
f(0),
\ee
It follows that $n=1$ always corresponds to $\alpha=1/2$. This is familiar
from the usual DGP result. We stress that all of these statements depend only on the leading
behavior of the spectral density as $s \rightarrow 0$ and so are independent of other details of the set-up.

\section{Cascading Scalars}
\label{scalar}

In this Section we review the recently proposed cascading mechanism~\cite{oriol} to regularize physics on a codimension-two brane, and generalize it to arbitrary codimensions.
For simplicity we focus here on a toy model of a scalar field, leaving the generalization to gravity to Secs.~\ref{gravSing} and~\ref{gravreg}.
The idea is to embed the codimension-two object on a codimension-one brane, each with their own intrinsic kinetic terms.
(In the gravitational case, these are intrinsic Einstein-Hilbert terms, as in the DGP model.) The resulting force law between test
particles on the codimension-two brane is therefore 4D at short distances, then effectively 5D over some
range of scales, and finally 6D at large distances. We
also show how this mechanism is completely generalizable to
arbitrary codimensions.

In what follows, we use the notation that capital Latin indices run over all the $D$ dimensions, $A,B = 0,1,\ldots, (D-1)$,
small Latin indices run over the $n=(D-4)$ extra dimensions, $a,b = 4,\ldots,(D-1)$,
while greek indices span over our four dimensions $\mu,\nu = 0,\ldots
3$. When working in $D=6$ dimensions, we will use the additional
notation that greek indices $\alpha, \beta$ span over the five
dimensions of a codimension-one brane. $M_d$ designates the
$d$-dimensional ``Planck" scale.

\subsection{Codimension-two case}
\label{CascScalarField}

In the codimension-two case,
the simplest construction is to consider a $(3+1)$-brane embedded on a $(4+1)$-brane, which itself is a codimension-one object in a 6D bulk.
Denoting the 6th dimension coordinate by $z$, and the 5th by $y$, the $(4+1)$-brane is taken to be at $z=0$, while the $(3+1)$-brane on its worldvolume is again at $y=z=0$. A $\mathbb{Z}_2$ symmetry is assumed across both branes, so the
$y$ and $z$-coordinate range from $0$ to $\infty$.
Unlike the pure codimension-two case, here we will find that the propagator is finite when evaluated on the codimension-two brane.
That is, the embedding on a codimension-one object removes the usual UV divergences of higher-codimension brane theories,~\cite{gw,claudia,geroch}.

The full action is given by
\be
S = -\frac{M_6^4}{2}\int d^6 x \, \partial_A \Phi\partial^A \Phi
-\frac{M_5^3}{2}\int d^5 x \, \partial_\alpha \Phi\partial^\alpha \Phi
-\frac{M_4^2}{2}\int d^4 x \, \partial_\mu \Phi\partial^\mu \Phi\,,
\label{S6da}
\ee
where the scalar field $\Phi$ is dimensionless.
The induced propagator between sources on the brane is usually calculated using the boundary effective action approach --- see, {\it e.g.},~\cite{lpr}.
Here we follow a different path, by treating the brane action as a localized coupling for the bulk free theory~\cite{gw,claudia}:
\be
S = -\frac{M_6^4}{2}\int d^6 x \, \partial_A \Phi\partial^A \Phi -  \int d^6 x \left\{\mathcal{L}_{\rm{coupling}}^{\rm{cod}1} + \mathcal{L}_{\rm{coupling}}^{\rm{cod}2}\right\}\,,
\label{S6dc}
\ee
with
\be
\mathcal{L}_{\rm{coupling}}^{\rm{cod}1} = \frac{M_5^3}{2} \delta(z)
\, \partial_\alpha \Phi \partial^\alpha \Phi\;; \qquad \mathcal{L}_{\rm{coupling}}^{\rm{cod}2}=\frac{M_4^2}{2} \delta(z)\delta(y)
\, \partial_\mu \Phi \partial^\mu \Phi\,.
\label{S6db}
\ee
This corresponds to a coupling $\lambda_1=-M_5^3 (q_1^2+p^2)$ on the codimension-one brane
($z=0$), where $q_1$ is the momentum in the $y$ direction, $p^2\equiv \eta^{\mu\nu}p_\mu p_\nu$ and $p_{\mu}$ the 4-momentum in the codimension-two brane directions $x^{\mu}$, along with a coupling $\lambda_2=-M_4^2 p^2$ localized on the codimension-two brane ($z=y=0$).

The free propagator associated with the bulk action is simply
\bea
\mathcal{G}_0(p_\mu,q_1; z,z')=\frac{1}{M_6^4}\int\frac{d q}{\pi} \frac{e^{i q(z-z')}}{p^2+q_1^2+q^2}\,,
\eea
where we work in the mixed representation, {\it i.e.}, in real space along $z$ and in momentum space along the  $y$ and $x^\mu$ directions, and $q$ is the momentum conjugate to $z$.
When both legs are taken on the codimension-one brane ($z=z'=0$), this propagator reduces to
\bea
\mathcal{G}_0(p_\mu,q_1)\equiv\mathcal{G}_0(p_\mu,q_1;0,0)=\frac{1}{M_6^4}\int\frac{d q}{\pi} \frac{1}{p^2+q_1^2+q^2} = \frac{1}{M_6^4\sqrt{p^2+q_1^2}}\,.
\label{dgp1}
\eea
Notice that in this scenario, due to the $\mathbb{Z}_2$ symmetry imposed across both the brane,
only half of the real plane for $z$ is considered. This explains the normalization taken for $q$.

\vspace{15pt}{\it \hspace{-15pt}$i$) Coupling on the codimension-one brane}\newline\indent

This free propagator is modified by the interaction term on the codimension-one brane.
These corrections are symbolically represented in Fig.~\ref{Coupling1}, where
in $n=6$ dimensions the coupling constant, $\lambda_1=-M_5^3 (p^2+q_1^2)$, is read off from
$\mathcal{L}_{\rm{coupling}}^{\rm cod1}$. Summing up these diagrams, we obtain the modified
Green's function $\mathcal{G}_1$ on the brane:
\bea
\label{DGPprop}
\mathcal{G}_1(p_\mu,q_1)&=&\mathcal{G}_0(p_\mu,q_1)+\lambda_1\mathcal{G}_0(p_\mu,q_1)^2
\(1+\lambda_1 \mathcal{G}_0(p_\mu,q_1)+\(\lambda_1 \mathcal{G}_0(p_\mu,q_1)\)^2+\cdots\) \nn \\
&=&\frac{\mathcal{G}_0(p_\mu,q_1)}{1+M_5^3 \(p^2+q_1^2\)\mathcal{G}_0(p_\mu,q_1)}
= \frac{1}{M_5^3}\frac{1}{p^2+q_1^2 + m_6\sqrt{p^2+q_1^2}}\,,
\eea
which is recognized as the usual DGP propagator, with 5D to 6D crossover scale
\be
m_6\equiv \frac{M_6^4}{M_5^3}\,.
\label{m6}
\ee
To proceed further it is useful to transform back to real space in $y$, so that
\ba
\mathcal G_1(p_\mu;y,y')=\frac{1}{M_5^3}\int \frac{d q_1}{\pi}
\frac{e^{i q_1 (y-y')}}{p^2+q_1^2 + m_6
\sqrt{p^2+q_1^2}}\,.
\ea
On the codimension-two brane, we therefore have
\be
\mathcal G_1(p_\mu) \equiv \mathcal G_1(p_\mu;0,0)
=\left\{\begin{array}{cl}
\frac{4}{\pi
M_5^3}\frac{1}{\sqrt{m_6^2-p^2}} \tanh^{-1}
\(\sqrt{\frac{m_6-p}{m_6+p}}\)\hspace{20pt}&\text{for}\hspace{10pt}
p<m_6 \\ \\
\frac{4}{\pi
M_5^3}\frac{1}{\sqrt{p^2-m_6^2}} \tan^{-1}
\(\sqrt{\frac{p-m_6}{p+m_6}}\)\hspace{20pt}&\text{for}\hspace{10pt}
p>m_6 \,.
\end{array}\right.
\ee

Notice that this integral is only finite as long as $M_5$ is
non-zero. Physically, the codimension-one brane plays the role of a
regulator without which the Green's function on the codimension-two
brane would be ill-defined. In particular, in the limit $M_5\rightarrow 0$ one
recovers the logarithmic divergence, $\mathcal{G}_1(p_\mu)\sim \int_0^{\Lambda} d q_1
\frac{1}{M_6^4 \sqrt{p^2+q_1^2}}\sim \log (\Lambda/p)$,
characteristic of codimension-two branes \cite{geroch}. Thus including a kinetic term on the codimension-one brane is crucial in order to avoid any singularity, making
the scalar field well-defined on the codimension-two brane. In the
gravitational case, the presence of kinetic terms on the codimension-two
brane will also be sufficient to avoid a singularity, as we will see in Sec.~\ref{gravSing}.

\vspace{15pt}{\it \hspace{-15pt}$ii$) Coupling on the codimension-two brane}\newline\indent
The same technique can be applied once more to take into
account the coupling $\mathcal{L}_{\rm{coupling}}^{\rm{cod}2}$ localized on the codimension-two
brane. The diagrams contributing to the new Green's function are as shown in Fig.~\ref{Coupling1}, with coupling
$\lambda_2 = - M_4^2 p^2$ localized at $(y,z)=(0,0)$. Using the same
summation, we therefore obtain the final Green's function on the
codimension-two brane
\be
\label{PropCod2}
\mathcal G_2(p_\mu)=\frac{\mathcal G_1(p_\mu)}{1+M_4^2\, p^2\, \mathcal G_1(p_\mu)}
= \frac{1}{M_4^2} \frac{1}{p^2 + g(p^2)}\,,
\ee
where for future reference we have defined
\ba
g(p^2) \equiv  \left\{\begin{array}{c}
\frac{\pi m_5}{4}\frac{\sqrt{m_6^2-p^2}}{\tanh^{-1}\(\sqrt{\frac{m_6-p}{m_6+p}}\)} \hspace{20pt}\text{for}\hspace{10pt}p<m_6\\ \\
\frac{\pi
m_5}{4}\frac{\sqrt{p^2-m_6^2}}{\tan^{-1}\(\sqrt{\frac{p-m_6}{p+m_6}}\)}
\hspace{30pt}\text{for}\hspace{10pt}p>m_6\,,
\end{array}\right.
\label{gp}
\ea
with $m_5$ the 4D to 5D cross-over scale,
\ba
m_5=\frac{M_5^3}{M_4^2}\,.
\label{m5}
\ea
Once again, in the absence of the original codimension-one brane,
corresponding to setting $M_5=0$ or, equivalently, $m_6\gg p$, the propagator on the brane would be ill-defined:
\bea
\mathcal G_2(p_\mu)\stackrel{p\ll m_6}{\longrightarrow} \frac{1}{2M_6^4}\log{p/m_6} \,.
\label{loglimit}
\eea
Thus the codimension-one brane plays the role of a regulator,
allowing for an arbitrary source to be localized on the
codimension-two brane.

We can also notice that we recover the DGP limit as $m_6\rightarrow 0$,
\ie when the field $\Phi$ effectively only propagates in five dimensions:
\bea
\mathcal G_2(p_\mu)\stackrel{p\gg m_6}{\longrightarrow} \frac{1}{M_4^2}\frac{1}{p^2+ m_5 p}
\,.
\eea

The previous example therefore exhibits a ``cascading" behavior,
interpolating from a four-dimensional behavior at short distances,
$p\gg m_5$, to a five-dimensional one for intermediate scales, $m_5\gg p \gg m_6$,
and finally a six-dimensional behavior at large distances, $p\ll m_6$. If $m_5>m_6$ then
there is a direct transition from 4D to 6D at a scale $p \sim \sqrt{m_5m_6}$.

\subsection{General Codimension-$n$}

\begin{figure}[]
\begin{picture}(150,50)(85,40)\thicklines
\put (-40,52){\makebox(0,0){$\mathcal{G}_m(p)$}}
\put (5,50){\circle*{2}}
\put (5,50){\line(1,0){45}}
\put (27.5,50){\circle*{10}}
\put (50,50){\circle*{2}}
\put (-10,49.5){\makebox(0,0){$=$}}
\put (65,49.5){\makebox(0,0){$=$}}
\put (80,50){\circle*{2}}
\put (80,50){\line(1,0){45}}
\put (125,50){\circle*{2}}
\put (140,49.5){\makebox(0,0){$+$}}
\put (155,50){\circle*{2}}
\put (155,50){\line(1,0){60}}
\put (215,50){\circle*{2}}
\put (185,50){\circle*{4}}
\put (185,40){\makebox(0,0){$\lambda_m$}}
\put (230,50){\makebox(0,0){$+$}}
\put (245,50){\circle*{2}}
\put (245,50){\line(1,0){90}}
\put (335,50){\circle*{2}}
\put (275,50){\circle*{4}}
\put (305,50){\circle*{4}}
\put (275,40){\makebox(0,0){$\lambda_m$}}
\put (305,40){\makebox(0,0){$\lambda_m$}}
\put (350,51){\makebox(0,0){$+$}}
\put (370,50){\makebox(0,0){$\cdots$}}
\normalsize
\end{picture}\\
\caption{Coupling corrections to the two-point function.
The Green's function on the codimension-$m$ brane is related to that
on the codimension-$(m-1)$, $\mathcal{G}_{m-1}(p)$.
$\lambda_m$ is the coupling  associated with
$\mathcal{L}_{\rm{coupling}}^{\rm{cod-m}}$:
$\lambda_m=-M_{4+n-m}^{2+n-m}\(p^2+q_1^2+\cdots+q_{n-m}^2\)$.}
\label{Coupling1}
\end{figure}

As can already been inferred at this point, this method can easily be
generalized to any higher codimension. In particular for a
codimension-$n$ brane, the full action will be given by
\be
S = - \frac{M_{n+4}^{n+2}}{2} \int d^{n+4} x \, \partial_M \Phi\partial^M\Phi
- \frac{M_{n+3}^{n+3}}{2} \int d^{n+3} x \, \partial_\alpha \Phi\partial^\alpha\Phi
-\cdots
- \frac{M_4^2}{2} \int d^4 x \partial_\mu\Phi\partial^\mu\Phi\,,
\ee
and a coupling $\lambda_m=-M_{4+n-m}^{2+n-m}(p^2+q_1^2+\cdots+q_{n-m}^2)$
will be present on each codimension-$m$ brane ($0 \le m \le n-1$). Here $p_\mu$ is the momentum along the $(3+1)$-brane and
$q_m$ the momentum along the $m^{\text{th}}$ extra
dimension. We denote by  $\mathcal{G}_{0}$ the free
$(4+n)$-dimensional Green's function
\ba
\mathcal{G}_0(p_\mu;q_1,\cdots,q_{n-1})=\frac{1}{M_{n+4}^{n+2}\sqrt{p^2+q_1^2+\cdots+
q_{n-1}^2}}\,,
\ea
evaluated on the codimension-one brane $\mathcal{G}_0\equiv\mathcal{G}_0(z^n=z'{}^n=0)$, and by $\mathcal{G}_{m}$ the
Green's function evaluated on the codimension-$m$ brane that takes into account
the couplings $\lambda_1,\cdots, \lambda_m$. This Green's
function is thus related to that on the codimension-$(m-1)$, $\mathcal{G}_{m-1}$,  by
\ba
&&\hspace{-20pt}\mathcal{G}_{m}(p_\mu;q_1,\cdots,q_{n-m-1})=\int \frac{d q_{n-m}}{\pi}
\frac{\mathcal{G}_{m-1}(p_\mu;q_1,\cdots,q_{n-m})}{1-\lambda_{m}\,
\mathcal{G}_{m-1}(p_\mu; q_1,\cdots,q_{n-m})}\,\nn\\ \nn\\
&&\hspace{50pt}=\int \frac{d q_{n-m}}{\pi}
\frac{\mathcal{G}_{m-1}(p_\mu;q_1,\cdots,q_{n-m})}{1+M_{4+n-m}^{2+n-m}(p^2+q_2^2+\cdots+q_{n-m}^2)\,
\mathcal{G}_{m-1}(p_\mu; q_1,\cdots,q_{n-m})}\,.
\ea
It follows that the field propagator projected on the codimension-$m$ $(3+1)$-brane is simply
given by
\ba
\mathcal G_{n}(p_\mu)=\frac{\mathcal G_{n-1}(p_\mu)}{1+M_4^2 p^2 \mathcal G_{n-1} (p_\mu)}\,,
\ea
 with $ \mathcal G_{n-1}$ satisfying the previous recursive
relation.
This gives rise to an arbitrary dimension cascading setup where the
field progressively switches from a four-dimensional to a $(4+n)$-dimensional behaviour as larger distances are considered.

\section{Cascading Gravity} \label{gravSing}

We would like to generalize this scalar field toy model to the
gravitational case, where brane induced kinetic terms are now
replaced by Einstein-Hilbert terms.
In this section, we will focus on the codimension-two case,
although the generalization to any codimension is straightforward.
In the scalar field toy-model, the presence of a kinetic term on the codimension-one brane effectively
regularizes the induced codimension-two propagator. The situation for gravity,
while similar, contains some subtleties.

In the six-dimensional scenario, the straightforward generalization of~(\ref{S6da}) to gravity is
\ba
\label{6dthin}
S=\frac{M_6^4}{2}\int d^6x \sqrt{-g_6} R_6
+\frac{M_5^3}{2}\int d^5x \sqrt{-g_5} R_5
+\frac{M_4^2}{2}\int d^4x \sqrt{-g_4} R_4\,.
\ea
In this case, the tensor modes behave precisely as the scalar field of Sec.~\ref{scalar}, as expected.
In particular, its induced propagator in four dimensions is made finite by the embedding on the codimension-one brane.
However, a key difference, as we will see, is that one of the scalar modes propagates a ghost~\cite{gigashif,sergei}.

We first present in what follows
the aforementioned pathology associated with~(\ref{6dthin}).
The detailed derivation of the gravitational exchange amplitude is
presented in Appendix~\ref{AppNaive}. We summarize here the main
result, mainly focusing on the behavior of the scalar modes, since
they are the ones that have a ghost-like behavior at high energies.

Working around flat space-time, the perturbations
may be decomposed into a scalar part and a four-dimensional transverse and traceless
(TT) part,
\ba
\label{4dTT}
h_{\mu \nu} &=& h_{\mu\nu}^{TT}
+ \pi  \eta_{\mu\nu}
+{\rm gauge\, terms}\,.
\ea
The tensor modes only couple to the conserved
and traceless part of the source, \ie to
$\Sigma_{\mu\nu}=T_{\mu\nu}^{(4)}-\frac{1}{3}T^{(4)}\eta_{\mu\nu}+\frac{\partial_\mu\partial_\nu}{3\Box_4}T^{(4)}$,
and their propagator is precisely that of the scalar field toy model
of section Sec.~\ref{scalar}. On the codimension-one brane, their
equation of motion is governed by
\ba
-\frac{M_5^3}{2}\left[\Box_5-m_6\sqrt{-\Box_5}\right] h_{\mu\nu}^{TT}=\delta(y)\left(\Sigma_{\mu\nu} +\frac{M_4^2}{2}\Box_4\,  h_{\mu\nu}^{TT}
\right)\,.
\ea
Their propagator is given in~\eqref{PropCod2}, and we obtain
\ba
\label{prop_hTT}
h_{\mu\nu}^{TT}=2\mathcal{G}_2
(p_\mu)\Sigma_{\mu\nu}=\frac{2}{M_4^2}\frac{1}{-\Box_4+g(-\Box_4)}\Sigma_{\mu\nu}\,.
\ea
Notice that since these modes behave identically to the scalar field
toy-model, they are completely regularized by the codimension-one brane.

The scalar mode, $\pi$, is also regularized by the codimension-one
brane,
\ba
-\frac{M_5^3}{2}\left[\Box_5-m_6\sqrt{-\Box_5}\right] \pi=
\frac{1}{12} \, \delta(y) T^{(5)}
= \frac{1}{12} \, \delta(y)\left(T^{(4)} -3M_4^2 \Box_4\pi\right)\,,
\label{Psieom}
\ea
and so will remain finite on the codimension-two brane, but is plagued
with a much worse pathology: $\pi$ has the wrong sign kinetic term
on the brane in the UV. Indeed in that regime, the left-hand side
of~\eqref{Psieom} is negligible, and $\pi$ satisfies $\pi = 1/ (3M_4^2 \Box_4) T
$, while in the IR the  left-hand side dominates and $\pi$ exhibits a 6D behavior
and couples to $- T$. Its kinetic term hence changes sign, signaling the appearence of a
ghost. In analogy with the transverse, traceless case, its solution is
\be
\label{prop_Psi}
\pi = \frac{2}{M_4^2}\frac{1}{\frac 12 \Box_4+g(-\Box_4)}\cdot \frac{1}{12}  T^{(4)}\,,
\ee
and its propagator is thus not positive definite. Notice that for
the tensor modes, the $\Box_4$ appears in \eqref{prop_hTT} with a
negative sign while it comes in with a positive sign for the scalar
mode \eqref{prop_Psi}.

The ghost manifests itself in the gravitational exchange amplitude ${\cal A}$ between two conserved source
$T_{\mu\nu}$ and $T'_{\mu\nu}$ on the brane:
\ba
\mathcal A &\sim&\int \mathrm{d}^4x \left( h_{\mu\nu}^{(4)TT} + \pi \eta_{\mu
\nu}\right)T'^{\mu\nu}\nn\\
\label{4dAmplitudeGhost}
&\sim&\frac{2}{M_4^2}\int \mathrm{d}^4x \;
T^{\mu\nu}\left\{\frac{1}{-\Box_4+g(-\Box_4)}\left(T'_{\mu\nu}-\frac
13 \eta_{\mu\nu}T'\right)
+\frac{1}{12}\;\frac{1}{\frac 12 \Box_4+g(-\Box_4)} \eta_{\mu\nu}T'\right\}\,.
\ea
The tensor structure is recognized as the sum of a massive spin-2 contribution, with the famous $1/3$ coefficient, plus a conformally-coupled
scalar. In the IR, corresponding to $-\Box_4 \ll g(-\Box_4)$, the overall coefficient of the $TT'$ term asymptotes to $-1/4$, as expected from the six-dimensional
behavior of the force law. In the UV, however, the overall $TT'$
coefficient tends to $-1/2$ which indicates a pathology.
In that regime, the scalar amplitude is indeed {\it negative}, indicative of a ghost.

The presence of this ghost is not specific to our cascading framework and appears
in general codimension-two and higher scenarios~\cite{gigashif,sergei}. To remove it we can modify the operators localised on the 3-brane. In \cite{oriol} it was argued that adding a tension to the codimension-two brane removes the ghost providing the tension satisfies $T_0 > \frac{2}{3}M_4^2m_6^2$. However, this tension is also necessarily bounded from above by $T_0 \le 2\pi M_6^4=2\pi m_5 m_6 M_4^2$, due to the normal requirement that the deficit angle of a codimension-two object must be less that $2\pi$ for the induced metric to remain flat.

In what follows we shall deal with an alternative way to remove the ghost, which does not require the introduction of a tension on the brane, but achieves the same effect of adding additional kinetic terms for $\pi$ from brane localized kinetic operators.

\section{Brane Regularizations}
\label{codnreg}

As mentioned in the previous section, the presence of the ghost
$\pi$ is generic to any codimension-two and higher
framework~\cite{gigashif,sergei}.
One can indeed recover the pure codimension-two scenario by
considering the limit where the five-dimensional Planck mass vanishes.
In that case  $m_6\gg \sqrt{-\Box_4}$ and the gravitational exchange amplitude is
still given by~\eqref{4dAmplitudeGhost}, with
\ba
g(-\Box_4^2)=\pi \frac{M_6^4}{M_4^2}\(\log
\frac{\Lambda}{\sqrt{-\Box_4}}\)^{-1}\,.
\ea
Here, $\Lambda$ is a UV cutoff that now needs to be
introduced since the geometry on the codimension-two brane is no
longer regularized by the codimension-one brane. In other words, $\Lambda$
represents the scale at which new physics is involved in the description of
the brane. Considering the codimension-one brane as a regulator, we
would have $\Lambda=2M_6^4/M_5^3$. We first focus on the regularization of the pure
codimension-$n$ scenario (no cascading), but the results are fairly
general and will be extended to the cascading framework in the next
section.

These inevitable divergences in higher codimension force us to regularize the branes. There are many ways to go about this, ranging from the elaborate construction of smooth topological defect models of the brane, to crudely imposing cutoffs. In the gravitational case it is crucially important that any chosen regularization scheme be gauge invariant.
In Appendix~\ref{appendix regularization}, we propose two
alternative regularization schemes that all lead to the same
result, as well as giving a description of how we understand the absence of a ghost by Kaluza-Klein reduction within the brane. We review these in the rest of this section after first discussing how we\
can always construct gauge invariant regularizations by means of introducing St{\"u}ckelberg fields.

\subsection{St{\"u}ckelberg Approach}
\label{gaugeinv}

A straightforward way to regulate branes is to replace them with `smeared branes' or `brane distributions'. Essentially we can think of a brane as being made up of N identical branes located at the same position. To smear them we spread the branes out so that we have some distribution.
To understand this, let us first consider the more straightforward example of a codimension-one brane. Imagine the brane is located at $\Phi(x,y)=0$, where $\Phi$ is some scalar function, then the action for the brane will be
\be
S_{\rm brane}=\int d^4 x \sqrt{-g_4} {\mathcal L}_M(g_{\mu\nu} ,\chi^i)\,,
\ee
where $\chi^i$ denotes matter fields localized on the brane, and $g_{\mu\nu}$ is the usual induced metric on the surface $\Phi=0$. This action is invariant under the full diffeomorphism group since 5D diffeomorphisms act as 4D diffeomorphisms on the surface of the brane.
We now smear this into a distribution $f(\epsilon)$ of branes localized at $\Phi(x,y)=\epsilon$, such that $\int_{-\infty}^{\infty} d \epsilon f(\epsilon)=1$, and define $g_{\epsilon,\mu\nu}$ as the associated induced metric. We thus have
\be
S^{\rm reg}_{\rm brane}=\int_{-\infty}^{\infty} d \epsilon f(\epsilon )\int d^4 x \sqrt{-g_{\epsilon}} {\mathcal L}_M(g_{\epsilon,\mu\nu} ,\chi^i)\,.
\ee
In general there is a 4D tensor ${\gamma}_{\mu\nu}(x,y)$ such that
\be
g_{\epsilon,\mu\nu}={\gamma}_{\mu\nu}(x,y) |_{\Phi=\epsilon}\,,
\ee
so we may write
\be
S^{\rm reg}_{\rm brane}=\int_{-\infty}^{\infty} d \epsilon f(\epsilon )\int d^4 x \int dy \partial_{y} \Phi \delta(\Phi-\epsilon)  \sqrt{-{\gamma}} {\mathcal L}_M({\gamma}_{\mu\nu} ,\chi^i)\,.
\ee
Performing the integral over $\epsilon$ we obtain
\be
S^{\rm reg}_{\rm brane}=\int d^4 x \int dy \partial_{y} \Phi f(\Phi) \sqrt{-{\gamma}} {\mathcal L}_M({\gamma}_{\mu\nu} ,\chi^i)\,.
\ee
Since the original action was gauge invariant, then so is this regularized form. However in practice it is most simple to understand this in the gauge in which $\Phi=y$ so that
\be
S^{\rm reg}_{\rm brane}=\int d^4 x \int dy f(y) \sqrt{-{g_{4}}} {\mathcal L}_M({g}_{\mu\nu} ,\chi^i)\,.
\ee

Following the same steps for codimension-two branes, defined as the locus of $\Phi_1=0$, $\Phi_2=0$ we have
\be
S^{\rm reg}_{\rm brane}=\int d^4 x \int dy dz \{\Phi_1,\Phi_2\} f(\Phi_1,\Phi_2) \sqrt{-{\gamma}} {\mathcal L}_M({\gamma}_{\mu\nu} ,\chi^i)\,,
\ee
where $\{\Phi_1,\Phi_2\} =\partial_y \Phi_1 \partial_z \Phi_2-\partial_z \Phi_1 \partial_y \Phi_2$. Again this is most straightforward to understand in the gauge where $\Phi_1=y,\Phi_2=z$ so that
\be
S^{\rm reg}_{\rm brane}=\int d^4 x \int dy dz f(y,z)\sqrt{-g_{4}} {\mathcal L}_M(g_{\mu\nu} ,\chi^i)\,,
\ee
or, in polar coordinates,
\be
S^{\rm reg}_{\rm brane}=\int d^4 x \int r dr d\theta \tilde{f}(r)\sqrt{-g_{4}} {\mathcal L}_M(g_{\mu\nu} ,\chi^i)\,.
\ee
Intuitively we can think of $\Phi_1$ and $\Phi_2$ as St{\"u}ckelberg fields corresponding to the gauge transformations which are spontaneously broken by the branes. In our calculations we mostly choose de Donder gauge in the bulk, which gives us sufficient gauge freedom to set the branes at a fixed position. This is equivalent to working in the gauge in which the St{\"u}ckelberg fields take the values $\Phi_1=y$ and $\Phi_2=z$.

Having established the gauge invariance of the smoothing procedure, in the rest of this section we present two alternatives regularization schemes that lead to the same low energy physics.

\subsection{Spherical Regularization} \label{sphreg}

In the first scheme, we replace the codimension-$n$ brane with a
codimension-one brane, wrapped around $(n-1)$ compact directions (see \cite{Kaloper:2007ap} for a very similar construction).
Expressing the Minkowski background in terms of the generalized
spherical coordinates along $n$ directions,
\be
ds^2_{D}=dr^2+r^2d\Omega^2_{n-1}+d^2x_{4}\,,
\ee
with $D=4+n$, and $r=0$ corresponds to a codimension-$n$ point.
If, instead, we choose to localize the brane at  $r=\Delta$, we are then
dealing with a codimension-one brane. And since this brane is now
$(3+n)$-dimensional, the corresponding intrinsic Einstein-Hilbert is
then  $R_{3+n}$ rather than $R_4$. As we can see in the Appendix,
this has dramatic consequences for the scalar degrees of freedom.
In particular, the induced metric perturbations sourced by an energy-momentum $T_{\mu\nu}$ on the brane is now
given by
\ba
h_{\mu \nu}=\mathcal{G}_\Delta(\Box_4)\(T_{\mu \nu}-\frac
1{D-2}T \eta_{\mu\nu}\)+\cdots
\label{rightanswer}
\ea
where the ellipses indicate total derivative, and the propagator ${\mathcal G}_\Delta
(\Box_4)$ tends to $\mathcal{G}_\Delta \sim -1/M_4^2 \Box_4$ in the
UV and to $(\sqrt{-\Box_4}/\Delta)^{2-n}/M_4^2$ in the IR. It should
be clear from this result that the resulting gravitational amplitude
is ghost-free. Notice, furthermore, that we obtain the same
coefficient for the $T T'$ amplitude at any energy scale. Strong
coupling effects will however modify this coefficient at high
energy, as we discuss in Sec.~\ref{strongcoupling}.

\subsection{Medium Model} \label{mediumstory}

In the second scheme, the codimension-$n$ brane is promoted to a
full $D=4+n$-dimensional object whose thickness $\Delta$ now plays
the role of the regularization scale. To realize this, we divide the entire space-time
into the two half-spaces, $r\leq \Delta$ and $r\geq \Delta$,
where the in-space models the blurred codimension-$n$ brane.

Since the brane is now $D$-dimensional, the natural curvature invariant to include on its world-volume is $R_D$
instead of $R_4$. In all generality this can have a different Planck scale from that of the ambient space-time. In other words,
the brane is thought of as a medium with gravitational permeability $\epsilon = M_4^{\; 2}/M_{D}^{\; D-2}$~\cite{massimo}. The gravitational part of the
action is therefore given by
\be
S =
\frac{M_{D}^{\; D-2}}{2}\int_{r>\Delta} d^D X\; \sqrt{- g_D} \; R_D+ \frac{\epsilon M_{D}^{\; D-2}}{2}\int_{r<\Delta} d^D X\; \sqrt{- g_D} \; R_D \; .
\label{mediumaction}
\ee
In this case the brane localized $R_4$ term arises naturally from the
$D$-dimensional Ricci curvature in the thin-brane limit $\Delta\rightarrow 0$.

Although the technical details in this case are slightly different, as shown in the
Appendix we still recover the same expression for the brane induced metric perturbations as~(\ref{rightanswer}),
giving rise to the same ghost-free amplitude.

\subsection{Effective Theory Approach}
\label{eff}

The spherical regularization and medium approach discussed above offer two independent smoothing procedures that lead to identical
ghost-free amplitudes.
Here we check this idea by deriving a low-energy effective action in the thin-brane limit, obtained by doing a KK reduction over the world-volume of
the regularized brane.

Indeed we find that either choice
leads to the same effective action as the thickness is taken to zero. Let us therefore consider the medium action~(\ref{mediumaction}) for concreteness.
Within $r<\Delta$, we take a metric ansatz (we can ignore the KK vectors since they are not sourced)
\be
ds^2_{r<\Delta} =  g_{\mu\nu}dx^\mu  dx^\nu + e^{2\phi} \left(dr^2 + r^2d\Omega^2_{D-5}\right)\,,
\ee
where $g_{\mu\nu}$ and $\phi$ are functions of $x^\mu$ only. In this approximation, we have $R_D = R_4 - 2(D-4)\Box_4\phi - (D-4)(D-3)(\partial_\mu\phi)^2$.
Substituting into~(\ref{mediumaction}) and integrating over the radial and angular directions, we obtain, after integration by parts, the following effective action
\be
S = \frac{M_D^{D-2}}{2}\int_{M_{\rm out}}d^Dx\sqrt{-g_D}R_D + \frac{M_4^2}{2}\int d^4x\sqrt{-g_4}e^{n\phi}\left(R_4+(D-4)(D-5)(\partial\phi)^2\right) \,,
\label{SeffD}
\ee
where $M_4^2 \sim \epsilon M_D^{D-2}\Delta^n$ is the induced 4D Planck scale. As mentioned earlier, exactly the same effective action follows from
the spherical regularization.

The effective brane action therefore reduces to an $R_4$ DGP term, although with additional scalar-tensor couplings to the extra-dimensional modulus $\phi$.
Note that in the standard DGP case, corresponding to $D=5$, the kinetic term for $\phi$ vanishes, while the extra factor of $e^\phi$ in the measure can be set to unity by
working in Gaussian normal coordinates. In other words, the modification is pure gauge in this case, leaving the standard DGP framework unaltered by the regularization.

However, the scalar-tensor structure is crucial for higher-codimension branes --- the extra conformal coupling affects the $TT'$ part of the exchange amplitude, thereby curing the ghost.
As shown in detail in the Appendix, by following similar steps to those of Sec.~\ref{gravSing} we arrive at the tensor structure displayed in~(\ref{rightanswer}), with the $TT'$
coefficient appropriate for $D$-dimensional embedding space-time. In what follows we apply these regularization techniques to the cascading framework.

\section{Ghost-free Cascading Gravity}
\label{gravreg}

To proceed we apply the general approach of Sec.~\ref{codnreg} ---
first smoothing out the brane to finite thickness and identifying the correct low energy theory.
The resulting action has a four-dimensional DGP term, albeit with additional scalar-tensor
couplings involving the extra-dimensional metric components. For concreteness, we henceforth
focus on the codimension-two set-up; the generalization to higher codimension is straightforward.

Remarkably, these new scalar-tensor couplings eradicate the ghost, leaving an exchange amplitude that is manifestly free of instabilities.
The corresponding force law extrapolates between a six-dimensional ($\sim 1/r^4$) behavior in the IR to a
four-dimensional ($\sim 1/r^2$) behavior in the UV. Meanwhile the tensor structure is that of six-dimensional gravity on
all scales.

Thus we begin by smoothing out the codimension-two brane to have finite thickness. As argued in Sec.~\ref{codnreg}, it no longer make sense
to add an $R_4$ term since the worldvolume is now 5+1 dimensional. Instead, the appropriate brane-induced gravity terms are either an $R_5$ term
on the (codimension-one) boundary of the regularized brane, as in Sec.~\ref{sphreg}, or an $R_6$ term throughout its worldvolume, as in Sec.~\ref{mediumstory}.

As argued in Sec.~\ref{eff}, either choice leads to the same effective action~(\ref{SeffD}) as the thickness is taken to zero, which for $D=6$ reduces to
\be
S_{\rm eff}^{(4)} = \frac{M_4^2}{2}\int d^4x\sqrt{-g_4}e^{2\phi}\left(R_4 + 2(\partial\phi)^2\right) + S_{\rm matter}[g]\,.
\label{Smod}
\ee
Note that the matter action is insensitive to the regularization, as argued in Sec.~\ref{gaugeinv}.
Thus the effect of regularization amounts to modifying the brane gravitational action to a scalar -tensor theory.

Conservation of stress energy requires that the variation of this action with respect to $\phi$ vanishes, which, at the linearized level, gives the condition
\be
R_4 = 2\Box_4\phi\,.
\label{R4phi}
\ee
Once again we can expand the 4D metric into scalar and TT components as in~(\ref{4dTT}). The TT part, $h_{\mu\nu}^{\rm TT}$, is by definition oblivious
to the trace part of the brane source, and is hence unaffected by the modified form of the brane effective action. The scalar-tensor structure does, however, affect the trace mode $\pi$.
Indeed, the trace of the effective source from~(\ref{Smod}) is
\be
T_{\rm eff} = \delta(y) \left\{M_4^2(R_4 - 6\Box_4\phi) + T^{(4)}\right\} = \delta(y) \left\{-2M_4^2R_4 + T^{(4)}\right\} \,.
\label{Teff}
\ee
Furthermore, in terms of the decomposition~(\ref{4dTT}) we have $R_4 = -3\Box_4\pi$. Hence,~(\ref{Psieom}) becomes
\be
-\frac{M_5^3}{2}\left[\Box_5-m_6\sqrt{-\Box_5}\right] \pi= \frac{1}{12} \, \delta(y)T + \frac{M_4^2}{2} \delta(y)\Box_4\pi\,.
\label{Psieommod}
\ee

Note that the kinetic operator now coincides with that in the TT equation. In particular, the brane-induced kinetic term for $\pi$ is healthy, which traces back to
the additional $\Box_4\phi$ term in~(\ref{Teff}) flipping the sign of $R_4$. The solution for $\pi$,
\be
\pi = \frac{2}{M_4^2}\frac{1}{-\Box_4+g(-\Box_4)} \cdot \frac{1}{12}T^{(4)}\,,
\ee
is therefore exactly of the same form as its TT cousin.

Thus the one-particle exchange amplitude between two conserved sources $T_{\mu\nu}$ and $T'_{\mu\nu}$ on the codimension-two brane takes the simple form
\be
\mathcal A \sim \frac{1}{M_4^2}\int \mathrm{d}^4x \; T^{\mu\nu}\frac{1}{-\Box_4+g(-\Box_4)}\left(T'_{\mu\nu}-\frac 14 \eta_{\mu\nu}T'\right) \,,
\ee
displaying the tensor structure of six-dimensional gravity. This is manifestly ghost-free since the coefficient of the $TT'$ part satisfies $-1/4 = -1/3 + 1/12$,
corresponding to a massive spin-2 plus a healthy conformally-coupled scalar. Notice that the above tensor structure is such that a cosmological constant does
not gravitate at the linearized level. This is consistent with the fact that pure tension on a codimension-two brane leaves the induced geometry flat and generates
instead a deficit angle in the embedding space. Note, however, that this codimension-two mechanism for taming the backreaction of vacuum energy is different than
degravitation.

This tensor structure is of course different from the usual $1/2$ predicted by general relativity. However, as in standard DGP, we expect that non-linearities
decouple the extra scalar modes in the vicinity of astrophysical sources, so that the theory approximates Einstein gravity locally. In particular, in the limit $m_6 < m_5$ in which gravity
has a 5D regime at intermediate distances, we expect a double Vainshtein effect: as we approach a source from infinity, first $\pi$ then $h_{\mu\nu}^{\rm TT}$
become strongly coupled, while the $TT'$ coefficient goes successively from $1/4$ to $1/3$ to $1/2$. We will come back to this issue in Sec.~\ref{strongcoupling}.

\section{Curing the Ghost in the Decoupling Limit}
\label{decoupling}

It is enlightening to study the origin of the ghost in the decoupling limit of the theory, which corresponds to sending $M_5,M_6\rightarrow \infty$
while keeping the strong coupling scale $(M_6^{16}/M_5^9)^{1/7}$ fixed.
The bulk physics can be integrated out in this limit, resulting in an effective 5D description in terms of a metric $h_{\alpha\beta}$ and a scalar mode $\pi$ encoding
brane bending in the extra dimension.

At quadratic order the 5D Lagrangian is given by
\be
{\cal L}_5 = \frac{M_5^3}{4}\left\{ -h^{\alpha\beta}({\cal E}h)_{\alpha\beta} - 3\pi\left(-\Box_5 h^\gamma_{\;\gamma} + \partial_\alpha\partial_\beta h^{\alpha\beta}\right)\right\} + \delta(y){\cal L}_4\,,
\label{L5}
\ee
where $h_{\alpha\beta}$ is the induced metric on the codimension-one brane, and ${\cal L}_4$ describes the codimension-two brane.
Since $\pi$ is a scalar, ${\cal L}_4$ can in principle depend both on the induced metric $h_{\mu\nu}$, as well as on this scalar mode. Thus let us assume the
general form
\be
{\cal L}_4 = \frac{M_4^2}{4}\left\{ -h^{\mu\nu}({\cal E}h)_{\mu\nu} -\beta\pi \left(-\Box_4 h^\mu_{\;\mu} + \partial_\mu\partial_\nu h^{\mu\nu}\right) + 3\gamma \pi\Box_4\pi\right\}+ \frac{1}{2}h^{\mu\nu}T^{(4)}_{\mu\nu}\,,
\ee
where $\beta$ and $\gamma$ are constants. In particular, the naive DGP extension studied in Sec.~\ref{gravSing} corresponds to $\beta = \gamma = 0$.

Following Appendix A, we find that $\pi$ appears on-shell in the decomposition of the 4D metric as
\be
h_{\mu\nu} = h_{\mu\nu}^{TT} + \pi\eta_{\mu\nu} + {\rm gauge\; terms}\,.
\ee
This agrees with the earlier decomposition --- e.g.~(\ref{4dTT}) ---, hence the choice of notation. Moreover, it immediately follows that $\pi$ is related on-shell
to the induced scalar curvature by $R_4 = -3\Box_4\pi$. Using this fact, the equation of motion for $\pi$ is thus
\be
-\frac{M_5^3}{2}\Box_5\pi = \frac{1}{12}\delta(y)T^{(4)} + \frac{M_4^2}{4}\delta(y)\left(\beta + \gamma - 1\right) \Box_4\pi\,,
\label{pigen}
\ee
from which we infer the following structure for the one-particle exchange amplitude
\be
T^{\mu\nu}\frac{1}{-\Box_4+m_5\sqrt{-\Box_4}}\left(T'_{\mu\nu}-\frac 13 \eta_{\mu\nu}T' + \frac{1}{12}\frac{-\Box_4 + m_5\sqrt{-\Box_4}}{-\Box_4\frac{[\beta+\gamma-1]}{2} + m_5\sqrt{-\Box_4}} \eta_{\mu\nu}T'\right) \,.
\ee
Therefore, the coefficient of the trace part interpolates between $-1/4$ at the largest distances, corresponding to the tensor structure of 6D gravity, and
\be
-\frac{1}{3}\left(1 - \frac{1}{2[\beta + \gamma-1]}\right)
\ee
at the shortest distances.

For instance, in the case $\beta= \gamma = 0$ of Sec.~\ref{gravSing}, the above equation for $\pi$ agrees with~(\ref{Psieom}) in the $m_6\rightarrow 0$ limit ---
the induced wave operator has the wrong sign, indicative of the ghost. More generally, though, the $\pi$ mode has a healthy propagator whenever
\be
\beta + \gamma > 1\,,
\label{betagamma}
\ee
which therefore requires non-zero $\beta$ and/or $\gamma$.

Let us therefore revisit our regularization schemes under this light, in particular in the effective theory language of Sec.~\ref{gravreg}. Since $R_4 = -3\Box_4\pi$ on-shell, comparison
with~(\ref{R4phi}) implies that $2\phi = -3\pi$. Substituting this in the expansion of~(\ref{Smod}) to quadratic order, we find that our regularization prescriptions all correspond to
$\beta = 6$ and $\gamma = -3$. In this case~(\ref{pigen}) reduces to
\be
-\frac{M_5^3}{2}\Box_5\pi = \frac{1}{12}\delta(y)T^{(4)} + \frac{M_4^2}{2}\delta(y)\Box_4\pi\,,
\label{pigen2}
\ee
again in agreement with the $m_6\rightarrow 0$ limit of~(\ref{Psieommod}).

Thus we see that our different ways of smoothing the codimension-two brane are just one example of how the ghost can be cured. Of course
not every choice of $\beta$ and $\gamma$ can necessarily be embedded within a covariant 6D theory. Another example where this is possible, however,
is through the addition of a tension $\Lambda$ on the codimension-two brane~\cite{oriol}. Thanks to $\pi$ self-interactions in the non-linear extension of~(\ref{L5}),
the non-zero tension generates an effective kinetic term for $\pi$ localized on the brane, corresponding to
\be
\gamma  = \frac{3}{2}\frac{\Lambda}{m_6^2M_4^2}\,.
\ee
And since $\beta = 0$ in this case,~(\ref{betagamma}) yields a lower bound on the required tension.

\section{Strong Coupling Phenomenon}
\label{strongcoupling}

Since massive gravitons in Minkowski spacetime exhibit the famous vDVZ discontinuity~\cite{vDVZ}, due to the persistence of an additional scalar degree of freedom in the $m \rightarrow 0$ limit, any theory in which the graviton is a resonance must exhibit strong coupling of any additional scalar modes at some scale, known as the Vainshtein scale, so as not to be ruled out by standard tests of General Relativity. In the 6D cascading model this means that the factor of $-1/4$ must effectively be converted to $-1/2$ in the strong coupling region. Here there are two additional scalar degrees of freedom, which intuitively arise from the $h_{zz}$ and $h_{yy}$ components of the metric. The scalar arising from the $6D \rightarrow 5D$ transition is $\pi$, and we expect it to become strongly coupled according to the natural generalization of the standard DGP story generalized to 6D. The scale at which strong coupling occurs, {\it i.e.} the Vainshtein scale, can be determined from the equation of motion for $\pi$ in the decoupling limit, including the nonlinear terms for $\pi$
\be
-\Box_5 \pi +\frac{9}{32m_6^2}\((\partial_{\alpha}\partial_{\beta}\pi)^2- (\Box\pi)^2\) = \delta(y) \frac{1}{6M_5^3}\( T+M_4^2 R^{(4)} \)\,,
\ee
by comparing the magnitude of the terms nonlinear in $\pi$, with the linear ones. To determine this we may first solve the linear equation, whose formal solution for $y\ge 0$ is
\be
\pi(x^{\mu},y) = e^{-\sqrt{-\Box_4}y} \pi_0(x^{\mu})\,.
\ee
Here $\pi_0$ is determined from the boundary condition imposed by the delta function source, $\partial_y \pi=-\frac{1}{6M_5^3}\( T+M_4^2 R^{(4)} \)$, to be
\be
\pi_0=\frac{1}{6M_5^3\sqrt{-\Box_4}}\( T+M_4^2 R^{(4)} \)\,.
\ee
Since $R^{(4)}$ contains terms of order $\Box_4 \pi$, there are two regimes of interest. For distances $r \gg m_5$ we have $\pi_0 \sim \frac{1}{6M_5^3\sqrt{-\Box_4}}T$, whereas for $r \ll m_5$ we expect $\pi_0 \sim\frac{1}{M_4^2 \Box}T$.

From the equation of motion, we see that strong coupling kicks in when $\partial \partial \pi \sim (\partial \partial \pi)^2/m_6^2$. In terms of the source this is when
\ba
r \gg m_5  &\quad&\partial T \sim m_6^2 M_5^3\,, \\
r \ll m_5 &\quad&  T \sim m_6^2 M_4^2\,,
\ea
and so for a mass $M$ with $T\sim M/r^3$, the Vainshtein scale is given by
\ba
r \gg m_5  &\quad& R_V = {}^4\sqrt{\frac{M}{m_6^2 m_5 M_4^2}}\,, \\
r \ll m_5 &\quad&R_V = {}^3\sqrt{\frac{M}{m_6^2 M_4^2} }\,.
\ea

Once the $\pi$ mode has become strongly coupled, the effective 5D theory reduces to pure gravity coupled to a codimension-one brane with induced kinetic terms. In other words, it looks like the standard DGP model, and so we anticipate that the second additional scalar will become strongly coupled at the usual scale
\be
\tilde{R}_V = {}^3\sqrt{\frac{M}{m_5^2M_4^2}}\,.
\ee
For this argument to be consistent we require $R_V \ge \tilde{R}_V$. For distances $r \ll m_5$, this is automatically true provided $m_6 \le m_5$ --- a necessary condition for a cascade from 6D to 4D via 5D.

Although the details of the strong coupling transitions should be checked by a careful calculation of the effective two field $\pi$-Lagrangian, we can see at least qualitatively that there can be a well defined cascade of strong coupling scales provided $m_6 < m_5$. We should stress that there is no reason to doubt that the same mechanism will also occur for $m_6 > m_5$. However, it will be necessary to do more explicit calculations in this case as there may be some subtle interaction between the two scalar modes which is not accounted for in the above argument.

\acknowledgments

We would like to thank Gregory Gabadadze and especially Gia Dvali, Oriol Pujolas and Michele Redi for useful discussions. The work of
CdR, SH, JK, and AJT at the Perimeter Institute is supported in part by the Government of Canada through NSERC and by the Province of Ontario
through MRI.

\newpage

\appendix

\section{Cascading Gravity ---  A Naive Approach}
\label{AppNaive}

Starting from the 6D action \eqref{6dthin}, and considering
perturbations around a flat metric
\ba
ds^2=(\eta_{AB}+h_{AB})dx^A dx^B\,,
\ea
in de Donder gauge $\partial_A h^A_B=\frac{1}{2}\partial_B
h^A_A$, the 6D Einstein equations take the simple form
\ba
\label{6dEinthin}
M_6^4 G^{(6)}_{AB}=-\frac{M_6^4}{2}\Box_{6}\(h_{AB}-\frac 12 h^C_{\, C}\,\eta_{AB}\)
=-\delta(z)\(M_5^2 G^{(5)}_{\alpha \beta}-T_{\alpha \beta}^{(5)}\)\delta^\alpha_{\, A}\delta^\beta_{\, B}\,.
\ea
Here the five-dimensional stress-tensor, $T_{\alpha \beta}^{(5)}$, includes the
four-dimensional Einstein term, $G^{(4)}_{\mu\nu}$, as well as matter stress-energy
localized on the $(3+1)$-brane, $T^{(4)}_{\mu\nu}$:
\ba
T^{(5)}_{\alpha\beta}=\delta(y)\(-M_4^2
G^{(4)}_{\mu\nu}+T^{(4)}_{\mu\nu}\)\delta^\mu_{\, \alpha}\delta^\nu_{\,
\beta}\,.
\label{Tab5}
\ea

\subsection{Step 1: $6D\rightarrow 5D$}

Since there are no energy-momentum sources along the $z$ direction, the $(\alpha,z)$ and $(z,z)$ Einstein
equations allow us to set $h_{\alpha z}=0$ and $h_{zz}=h^\alpha_\alpha$, respectively. The induced
five-dimensional gauge choice is thus
$\partial_\alpha h^\alpha_{\, \beta}-\partial_\beta h^\alpha_{\,
\alpha}=0$,
in terms of which the five-dimensional Einstein tensor takes the form
\ba
G^{(5)}_{\alpha\beta}=-\frac12 \(\Box_5 h_{\alpha\beta}-\partial_\alpha \partial_\beta h^\gamma_{\,
\gamma}\)\,.
\ea
The remaining five-dimensional Einstein equations are then
\ba
-\frac{M_6^4}{2}\(\Box_6+\frac{\delta (z)}{m_6}\Box_5\)h_{\alpha
\beta}=\delta(z)\(T^{(5)}_{\alpha \beta}-\frac 14 T^{(5)}\,\eta_{\alpha\beta}\)
-\frac{\delta(z)}{2}M_5^3 \partial_\alpha \partial_\beta
h^\gamma_{\,\gamma}\,,
\ea
with $m_6$ defined in~(\ref{m6}). In particular, the trace part satisfies
\ba
\label{singmode2}
-\frac{M_6^4}{2}\, \Box_6 h^\gamma_{\,\gamma}&=&-\frac 14 \delta(z) T^{(5)}\,.
\ea
Note that, since $T^{(5)}\sim \delta(y)$, this mode only couples to the codimension-two source.
That is, it does not ``feel" the codimension-one brane, and its solution therefore suffers from logarithmic
divergences characteristic of codimension-two branes. However this is purely a gauge artifact as can be seen by going to the gauge defined by $h_{zM}=0$, where the metric is given by
\be
\tilde{h}_{\alpha\beta} = h_{\alpha\beta}^{(5d)TT}+\frac{|z|}{2M_6^4}\frac{\partial_{\alpha}\partial_{\beta}}{\Box_5}T^{(5)}\,.
\ee
In the above we have found it convenient to project out the 5D transverse and traceless variable,
\ba
h_{\alpha \beta}^{(5d)TT}=h_{\alpha
\beta}-\frac{\partial_\alpha\partial_\beta}{\Box_5}h^\gamma_{\;\gamma}\,,
\ea
which satisfies the decoupled and fully regularized equation
\be
-\frac{M_6^4}{2}\(\Box_6+\frac{\delta (z)}{m_6}\Box_5\)h_{\alpha \beta}^{(5d)TT}=\delta(z)\(T^{(5)}_{\alpha \beta}-\frac 14 T^{(5)}\,\eta_{\alpha\beta}+\frac{1}{4} \frac{\partial_{\alpha}\partial_{\beta}T^{(5)}}{\Box_5}\)\,.
\ee
This can be solved exactly following the $6D\rightarrow 5D$ reduction in the scalar toy model --- see~(\ref{CascScalarField}):
\ba
\label{5dEq}
-\frac{M_5^3}{2}\left[\Box_5-m_6\sqrt{-\Box_5}\right] h_{\alpha \beta}^{(5d)TT}
&=&\delta^\mu_\alpha\delta^\nu_\beta \left[T_{\mu\nu}^{(4)}-M_4^2 G_{\mu\nu}^{(4)}\right]\delta(y) \\
& & \;\;\;\; -\frac 14 \(\eta_{\alpha\beta}-\frac{\partial_\alpha\partial_\beta}{\Box_5}\)\left[T^{(4)}+M_4^2 R^{(4)}\right]\delta(y)\,,
\ea
where $h_{\alpha \beta}^{(5d)TT}$ now denotes the metric fluctuation evaluated on the codimension-one brane, and where we have substituted
for $T_{\alpha \beta}^{(5)}$ using~(\ref{Tab5}). Note that the derivative operator on the left-hand side is recognized as the inverse of
the scalar Green's function~(see Sec.~\ref{CascScalarField}).

\subsection{Step 2: $5D\rightarrow 4D$}
Writing $\pi = -\frac{\Box_5}{3\Box_4} h^{(5d)TT}_{yy}= \frac{\Box_5}{3\Box_4} \eta ^{\mu \nu}h_{\mu
\nu}^{(5d)TT}$, we can express the 4D transverse and traceless part
as
\be
h_{\mu \nu}^{(4d)TT} =h_{\mu \nu}^{(5d)TT}-  \pi
\eta_{\mu\nu}-\(3\frac{\Box_4}{\Box_5}-4\)\frac{1}{\Box_4}\partial_\mu\partial_\nu\pi\,,
\ee
giving the following expression for the four-dimensional Einstein
tensor
\be
G^{(4)}_{\mu\nu} =-\frac12\(\Box_4 h_{\mu \nu}^{(4d)TT} - 2 \Box_4 \(\eta_{\mu\nu}-\frac{\partial_\mu \partial_\nu}{\Box_4}\)\pi\)\,.
\ee
Meanwhile, the $(y,y)$ component of~\eqref{5dEq} yields an equation of motion for $\pi$:
\ba
-\frac{M_5^3}{2}\left[\Box_5-m_6\sqrt{-\Box_5}\right] \pi=\frac{1}{12}\delta(y)
\(T^{(4)}-3M_4^2\Box_4 \pi\)\,.
\label{psieom}
\ea
Putting everything together, the $(\mu \nu)$ component of~\eqref{5dEq} simplifies to
\ba
  -\frac{M_5^3}{2}\left[\Box_5-m_6\sqrt{-\Box_5}\right] h_{\mu\nu}^{(4d)TT}=\delta(y)\left(\Sigma_{\mu\nu} +\frac{M_4^2}{2}\Box_4\  h_{\mu\nu}^{(4d)TT} \right)\,,
\label{barheom}
\ea
where $\Sigma_{\mu\nu}=T_{\mu\nu}^{(4)}-\frac{1}{3}T^{(4)}\eta_{\mu\nu}+\frac{\partial_\mu\partial_\nu}{3\Box_4}T^{(4)}$ is the transverse and traceless
part of the matter stress tensor.

Equation~(\ref{barheom}) is the exact analogue of the scalar field equation derived from~(\ref{S6da}). Thus, as advocated, the transverse and traceless story
parallels the scalar toy model presented earlier. The solution is therefore given by
\be
 h_{\mu\nu}^{(4d)TT}=\frac{2}{M_4^2}\frac{1}{-\Box_4+g(-\Box_4)} \Sigma_{\mu\nu}\,,
\label{barhans}
\ee
where the function $g$ was introduced in~(\ref{gp}), and $h_{\mu\nu}^{(4d)TT}$ now denotes the induced metric fluctuation on the codimension-two brane.

Equation~\eqref{psieom} for the scalar mode $\pi$, on the other
hand, is slightly different as that for the tensor modes and gives
rise to the solution
\be
\label{solPsi}
\pi = -\frac{1}{3M_4^2}\frac{1}{-\Box_4-2g(-\Box_4)} T^{(4)}\,.
\ee
A key difference between~(\ref{barhans}) and~(\ref{solPsi}) is the
presence of the negative sign for $g(-\Box_4)$ in the
expression of $\pi$, giving rise to the presence of a
ghost as soon as $\Box_4\gg g(-\Box_4)$.

\section{Codimension-$n$ brane regularization}
\label{appendix regularization}

\subsection{Spherical Regularization}

Let us work in $D=(4+n)$ dimensions where $n$ is the codimension. Consider the action representing the codimension-$n$ version of the DGP model:
\be
S=\frac{M_D^{D-2}}{2}\int d^D x \sqrt{-g_D}R_D+\frac{M_4^2}{2}\int d^4 x \sqrt{-g_4}R_4+\int d^4x \sqrt{-g_4}{\mathcal
L}_m\,.
\ee
In this section we replace the codimension-$n$ brane with a codimension-one brane which is wrapped around
$(n-1)$ compact directions (see \cite{Kaloper:2007ap}).

In particular, let us assume that the bulk metric is given by Minkowski spacetime. It is useful to write this in the generalization of spherical or polar coordinates in the $n$ directions as
\be
ds^2_D=dr^2+r^2d\Omega^2_{n-1}+d^2x_{4}\,,
\ee
where by $d\Omega^2_{n-1}$ we understand the line element on a unit $(n-1)$-sphere. The point $r=0$ is a codimension-$n$ point where we assume the brane to be located. To regulate we replace the codimension-$n$ brane with a codimension-one brane located at $r=\Delta$ so that the induced metric on the codimension-one brane is
\be
ds^2_{\rm induced}=\Delta^2 d\Omega^2_{n-1}+d^2x_{4}\,.
\ee
In this way we can take our matter fields to be spread out over the codimension-one brane since in the limit $\Delta \rightarrow 0$ we are essentially performing a KK compactification and the dynamics on the brane will be dominated by the KK zero modes which are effectively living in $D$ dimensions. The KK reduction approach will be pursued in Sec.~\ref{justin}.
In particular this is true for energy on the brane $E \ll \Delta^{-1}$.
Strictly speaking we should address the stability of putting the
brane here, however it is likely easy to cook up a stabilization
mechanism using e.g. form fields. Nevertheless we can still address
whether this set-up has ghosts independent of this, since even an
unstable solution should be unitary.

Defining $\gamma^{-1} = \Delta^{n-1}V_{n-1}$ where $V_{n-1}$ is the volume of the $(n-1)$-sphere, the regulated action is
\be
S=\frac{M_D^{D-2}}{2}\int d^D x \sqrt{-g_D}R_D+\frac{\gamma M_4^{2}}{2}\int d^{D-1} x \sqrt{-g_{D-1}}R_{D-1}+\gamma \int d^{D-1}x \sqrt{-g_{D-1}}{\mathcal L}_M\,.
\ee
In practice we can neglect the KK excitations of the matter fields.  For $n\ge3$ we include a tension in the spherical directions to support the background solution for otherwise the extrinsic curvature will pick up a jump due to the nonzero curvature in the spherical directions. Since this is pure tension, however, it will not contribute in the perturbed equations, and its only role is to support the background solution.
Furthermore, the tension is needed for regularization purposes only and disappears in the thin-brane limit.

The perturbed equations of motion are thus given by
\be
M_D^{D-2}G^{(D)\, A}_{\phantom{(D)\, }\;B}+ \delta(r-\Delta) \gamma  M_4^{2}G^{(D-1)\, a}_{\phantom{(D-1)\, }\;b}\,  \delta^A_{\;a}\delta^b_{\;B}=\gamma \delta(r-\Delta) T^a_{\;b}\,  \delta^A_{\;a}\delta^b_{\;B}\,,
\ee
where $A,B,\ldots$ span as usual over all $D$ dimensions,
whereas $a$ spans over all but the transverse coordinate $r$.
It should be evident that this is invariant under the linearized version of $D$-dimensional diffeomorphisms.

The reduction from $D$ dimensions to $(D-1)$ is now familiar.
We again choose the $D$-dimensional de Donder gauge
$\nabla_A\left(h^A_{\;B}-\frac{1}{2}\delta^A_{\;B}h^C_{\;C}\right)=0\,.$
And since there are no sources in the $(r,A)$ directions we infer as
previously, $h^r_{\;a}=0$ and $h^r_{\;r}-\frac{1}{2}h^C_{\;C} =0\,,$
which implies $h^r_{\;r}=h^a_{\;a}$. The de Donder gauge condition
is then $\nabla_a\left(h^a_{\;b}-\delta^a_{\;b}h^c_{\;c}\right) =
0$, which implies $G^{(D-1)}_{ab} =-\frac{1}{2}
\left(\Box_{D-1}h_{ab}-\partial_a\partial_bh^c_{\;c}\right)\,.$
Since this is automatically traceless we infer that
\be
-\frac{1}{2}M_D^{D-2}(2-D)\Box_D h^c_c=\gamma \delta(r-\Delta)T\,,
\ee
and so by defining the traceless part of $h_{ab}$ as
\be
H_{ab}=h_{ab}-\frac{1}{\Box_{D-1}} \nabla_a\nabla_bh^c_{\;c}\,,
\ee
we have, after substituting in the trace,
\be
-\frac{1}{2}M_D^{D-2}\Box_D
H_{ab}-\frac{1}{2}\gamma \delta(r-\Delta)M_4^{2}\Box_{D-1}H_{ab}=\gamma \delta(r-\Delta) \left(
T_{ab}-\frac{1}{(D-2)}\eta_{ab}T\right)+\ldots\,,
\ee
where the ellipses indicate a total derivative, which are irrelevant when coupling to a conserved source.
In the limit $\Delta \rightarrow 0$ we have $\Box_{D-1} \rightarrow \Box_{4}$ according to the conventional KK prescription. Defining the
Green's function $\mathcal G$ as the solution of
\be
M_D^{D-2}\Box_D \mathcal G(x;x') +\gamma\delta(r-\Delta)M_4^{2} \Box_{4} \mathcal G(x;x') = \delta^{(D)}(x;x')\,,
\ee
we infer that the gravitational exchange amplitude is given by
\ba
\mathcal{A}&\sim& \int d^4x h_{\mu \nu}T'{}^{\mu\nu}\sim \int d^4x H_{\mu
\nu}T'{}^{\mu\nu}\notag\\
&\sim& \int d^4x\ \mathcal G_\Delta
(x^\mu)\left\{T_{\mu\nu}-\frac{1}{(D-2)}\eta_{\mu\nu}T\right\}T'{}^{\mu\nu}\,,
\ea
where $ \mathcal G_\Delta (x^\mu)\equiv \mathcal G(x^\mu, r=\Delta; 0,
r'=\Delta)$.
This is manifestly ghost free since we can write it as
\be
\mathcal{A}\sim \int d^4x \ \mathcal G_\Delta(x^\mu)
\left\{\left(T_{\mu\nu}-\frac{1}{3}\eta_{\mu\nu}T\right)T'{}^{\mu\nu}+C \, T\, T'\right\}\,,
\ee
where $C=\frac13-\frac{1}{D-2} \ge 0$ for $D \ge 5$.

\subsection{Medium Model for a Higher-Codimension Brane}

In this regularization, we consider the two half-spaces
\begin{equation}
M_{\rm in} = \left\{X\mid r \le \Delta\right\} \; , \hspace{0.5cm}
M_{\rm out} = \left\{X\mid r \ge \Delta\right\} \; ,
\end{equation}
with $r$ denoting the Euclidean distance in the transverse direction and $\Delta$ the
thickness of the in-space in the transverse directions. The in-space models the blurred
codimension-$n$ brane. We might think of it as a medium with gravitational permeability
$\epsilon = M_4^{\; 2}/M_{D}^{\; D-2}$. The action for this set-up is thus
\begin{equation}
\label{action}
S =
\frac{M_{D}^{\; D-2}}{2}\int_{{M_{{\rm out}}}} d^D X\; \sqrt{- g_D} \; R_D+ \frac{\epsilon M_{D}^{\; D-2}}{2}\int_{{M_{{\rm in}}}} d^D X\; \sqrt{- g_D} \; R_D \; ,
\end{equation}
which corresponds to integrating over a sharp profile $P(r)\equiv \epsilon \theta(\Delta-r) + \theta(r-\Delta)$ that describes the support of $\sqrt{- g_D} \; R_D$ in $M_{\rm in}$ and
$M_{\rm out}$.


Varying the above action gives rise to Einstein's equations in both half-spaces,
$M_{\rm in}$ and $M_{\rm out}$, along with the following boundary conditions
\begin{eqnarray}
    \label{bc}
\epsilon \; n^A J_A \mid_{\partial M}
&=&
n^A J_A \mid_{\partial M}
\; .
\end{eqnarray}
These are the analogue of the Israel jump conditions.
In the weak-field approximation, the current density through
the interior surface becomes $J^A = \partial^A h^B_B - \partial_B h^{AB}$.
Imposing the $4$-gauge conditions $\partial_a h^a_{\, \mu} = 0$,
the Einstein tensor in the $(4+n)$-split is given by
\begin{eqnarray}
({\mathcal E}_D h)_{\mu\nu}
\label{Einb}
&=&
-\frac{1}{2}\left(({\mathcal E}_4 h)_{\mu\nu} + \left(\partial_\mu \partial_\nu - \eta_{\mu\nu}\Box_4\right) h_n\right.
\nonumber \\
&&\left. + \Box_n \left[h_{\mu\nu}-\eta_{\mu\nu}\left(h_4 +
h_n\right)\right]
+ \eta_{\mu\nu} \partial^a \partial^b h_{ab}\right)
\; , \\
({\mathcal E}_D h)_{ab}
\label{Eint}
&=&
-\frac{1}{2}\left(\left(\Box_4 + \Box_n\right) h_{ab}
+\left[\partial_a \partial_b - \eta_{ab} \left(\Box_4 + \Box_n\right)\right]
\left(h_4 + h_n\right)\right.
\nonumber \\
&&\left.- \partial^c \partial_{
\left(a\right.
} h_{
\left.b\right)c
} + \eta_{ab} \partial^c\partial^e h_{c e}
- \partial^\mu \partial_{\left(a\right.} h_{\left.b\right)\mu} +
\eta_{ab} \partial^\mu\partial^\nu h_{\mu\nu}\right)
\; ,
\end{eqnarray}
where the traces are defined as follows
$h_4\equiv \eta^{\mu\nu}h_{\mu\nu}$ and $h_n\equiv \eta^{ab}h_{ab}$.
The mixed components of the Einstein tensor are not required.

We are interested in the polarization tensor of the induced weak field $h_{\mu\nu}$.
In the case of induced gravity, the integration theory of the weak field equations of motion requires us to first gauge shift to a new field $\Psi_{\mu\nu}$ that obeys a generalized
Fierz-Pauli equation after appropriate gauge fixing. The form of the gauge-shift is clear from
(\ref{Einb}):
\begin{equation}
h_{\mu\nu} = \Psi_{\mu\nu} + {\cal C} h_n \eta_{\mu\nu}
\; ,
\end{equation}
with ${\cal C}$ denoting an adjustable constant. Imposing further the $D-4$ gauge conditions $\partial_a h^a_{\,b} = -\partial_b h_n/2$, we find for ${\cal C}=-1/2$
\begin{equation}
\label{eomP}
({\mathcal E}_4 \Psi)_{\mu\nu} - m^2(\Box_n) \left(\Psi_{\mu\nu} - \Psi_4 \eta_{\mu\nu}\right)
= \frac{1}{M_D^{D-2}}T_{\mu\nu}
\; ,
\end{equation}
where we have allowed for an (induced) source term. This is indeed Fierz-Pauli theory for the shifted gauge field $\Psi_{\mu\nu}$ with a generalized mass term $m^2 = -\Box_n$ in
accordance with the usual Kaluza-Klein reduction.

Since only transverse degrees of freedom propagate in Fierz-Pauli theory, the trace
of the gauge field can be eliminated for a conserved source,
$\Psi_4 = 2T_4/3M_D^{D-2}\Box_n$, allowing us to solve (\ref{eomP}):
\begin{equation}
    \Psi_{\mu\nu}
    =
    -\frac{2}{M_D^{D-2}} \frac{1}{\Box_4 - m^2(\Box_n)}\left(T_{\mu\nu} - \frac{1}{3}\eta_{\mu\nu}T\right)\,.
    \ee
The boundary conditions (\ref{bc}) impose $m^2(\Box_n)=f(\Box_4)$.
However, $\Psi_{\mu\nu}$ depends on $h_n$. In order to eliminate $h_n$ we write down the equations of motion for $h_{ab}$:
\begin{equation}
    \label{eomij}
    \left(\Box_4 - m^2(\Box_n)\right) \left(h_{ab} + \frac{1}{2} h_n \eta_{ab}\right)
    - \partial^\mu\partial_{(a}h_{b)\mu}
    =
    \frac{2}{3M_D^{D-2}} \left(\eta_{ab} - \frac{\partial_a \partial_b}{-m^2(\Box_n)}\right) T
    \; .
\end{equation}
Tracing (\ref{eomij}) and using again $\partial^ah_{a\mu} = 0$ gives $h_n$ entirely in terms of the induced source term and the appropriate Green-function:
\begin{equation}
    \label{hn}
    h_n =  \frac{4}{3M_D^{D-2}} \; \frac{D-5}{D-2} \; \frac{T_4}{\Box_4 - m^2(\Box_n)} \;.
\end{equation}
Note that for $D=5$ the source term vanishes. For $D>5$ the solution (\ref{hn}) guarantees that $h_{\mu\nu}$ has the correct polarization tensor. Indeed,
\begin{equation}
\label{hMedium}
    h_{\mu\nu}
    =
    -\frac{2}{M_D^{D-2}} \frac{1}{\Box_4 - m^2(\Box_n)}\left(T_{\mu\nu} - \frac{1}{D-2}\eta_{\mu\nu}T\right)\,.
\end{equation}

\subsection{Dimensional Reduction in Zero thickness limit}
\label{justin}

As argued in Sec.~\ref{codnreg}, both regularization schemes described above lead to the same 4D effective brane action in the thin-brane limit.
To review, we perform a KK reduction within the world-volume of the thick brane by taking the ansatz $ds_D^2 = g_{\mu\nu}dx^\mu  dx^\nu + e^{2\phi} \left(dr^2 + r^2d\Omega^2_{D-5}\right)$.
The resulting effective action is given by
\be
S = \frac{M_D^{D-2}}{2}\int d^Dx\sqrt{-g_D}R_D + \frac{M_4^2}{2}\int d^4x\sqrt{-g_4}e^{n\phi}\left(R_4+(D-4)(D-5)(\partial\phi)^2\right) \,.
\ee
A key difference with the usual DGP Einstein-Hilbert term on the defect are the extra terms involving $\phi$. These will prove instrumental in getting the desired tensor structure for the exchange amplitude.

As previously, in de Donder gauge the Einstein equations take the form
\be
-\frac{M_D^{D-2}}{2}\Box_D\left(h^A_{\;B}-\frac{1}{2}\delta^A_{\;B}h^C_{\;C}\right) =
T^{A\;(6)}_{\;B}\,.
\ee
The $\phi$-dependent terms in the brane action contribute to the $(r,r)$ component of the stress tensor. Conservation of energy-momentum therefore requires this contribution to vanish,
which, at the linearized level, yields the condition
\be
R_4 = 2(D-5)\Box_4\phi\,.
\label{R4}
\ee
Meanwhile, the stress tensor components along the brane are given by
\be
T^{\mu\;(6)}_{\;\nu} = \delta^{(n)}(r)\left\{-M_4^2G^{\mu\;(4)}_{\;\nu}-(D-4)M_4^2\left(\delta^\mu_{\;\nu}\Box_4-\partial^\mu\partial_\nu\right)\phi+T^\mu_{\;\nu}\right\}\,.
\ee

Now the choice of gauge in the bulk is consistent with the gauge relations $\partial_\nu h^\nu_{\;\mu}  = 2\partial_\mu\phi$ and $h^\mu_{\;\mu} = 2(6-D)\phi$.
In particular, the linearized 4D Einstein tensor takes the form
\be
G_{\mu\nu}^{(4)} = -\frac{1}{2}\left(\Box_4h_{\mu\nu} +2\eta_{\mu\nu}(D-5)\Box_4\phi-2(D-4)\partial_\mu\partial_\nu\phi\right)\,,
\ee
whose trace is indeed consistent with~(\ref{R4}).

Substituting into the $(\mu,\nu)$ equations of motion, we find
\be
-\frac{1}{2}\left(M_D^{D-2}\Box_6 + \delta^{(n)}(r)M_4^2\Box_4\right) \left(h^{\mu}_{\;\nu}-2\delta^\mu_{\;\nu}\phi\right) = \delta^{(n)}(r) T^\mu_{\;\nu}\,.
\label{munueqn}
\ee
The scalar term can be eliminated by taking the trace,
\be
-\left(M_D^{D-2}\Box_6 + \delta^{(n)}(r)M_4^2\Box_4\right)\phi = -\frac{1}{D-2}\delta^{(n)}(r) T\,,
\ee
and substituting the result back into~(\ref{munueqn}) to obtain
\be
-\frac{1}{2}\left(M_D^{D-2}\Box_6 + \delta^{(n)}(r)M_4^2\Box_4\right) h^{\mu}_{\;\nu} = \delta^{(n)}(r) \left( T^\mu_{\;\nu} -\frac{1}{D-2}\delta^\mu_{\;\nu}T\right)\,.
\ee
This agrees with our earlier results.

\pagebreak

\end{document}